\documentclass[sigconf]{acmart}
\AtBeginDocument{%
  }

\setcopyright{acmlicensed}
\copyrightyear{2026}
\acmYear{2026}
\acmDOI{XXXXXXX.XXXXXXX}
\acmConference[EASE 2026]{The 30th International Conference on Evaluation and Assessment in Software Engineering}{9–12 June, 2026}{Glasgow, Scotland, United Kingdom}
\acmISBN{978-1-4503-XXXX-X/2018/06}

\usepackage{graphicx}%
\usepackage{multirow}%
\usepackage{amsmath}
\usepackage{amsthm}%
\usepackage{mathrsfs}%
\usepackage[title]{appendix}%
\usepackage{xcolor}%
\usepackage{textcomp}%
\usepackage{manyfoot}%
\usepackage{booktabs}%
\usepackage{algorithm}%
\usepackage{algorithmicx}%
\usepackage{algpseudocode}%
\usepackage{listings}%

\newcommand{\util}{\texttt{util} }
\newcommand{\utiln}{\texttt{util}}
\newcommand{\helper}{\texttt{helper} }

\newcommand{\test}{\texttt{test} }

\newcommand{\goal}{\textit{to help developers avoid creating unsafe and unused \util files when developing their projects}}

\newcommand{\rqone}{RQ 1. Usage}
\newcommand{\rqonev}{How are \util files being used and/or reused in mature open source projects?}

\newcommand{\rqoneone}{RQ 1.1 Prevalence}
\newcommand{\rqoneonev}{How many \util files are in mature open source projects?}

\newcommand{\rqonetwo}{RQ 1.2 Invocation}
\newcommand{\rqonetwov}{How often are the methods in \util files invoked?}

\newcommand{\rqonethree}{RQ 1.3 Renaming}
\newcommand{\rqonethreev}{Is the \util naming convention more likely to be adopted, abandoned, or oscillating over time?}

\newcommand{\rqtwo}{\textbf{RQ 2. Complexity} }
\newcommand{\rqtwov}{Are \util files more complex than non-\util files?}

\newcommand{\rqthree}{\textbf{RQ 3.} Collaboration}
\newcommand{\rqthreev}{Are \util files being developed as collaborative efforts?}

\newcommand{\rqfour}{\textbf{RQ 4.} Security}
\newcommand{\rqfourv}{How have \util files compared to non-\util files in terms of presence in vulnerability fixes?}

\newcommand{\rqfourone}{RQ 4.1 Risk Factor}
\newcommand{\rqfouronev}{Are \util files more likely to be in a vulnerability fix than non-\util files?}

\newcommand{\rqfourtwo}{RQ 4.2 Recidivism}
\newcommand{\rqfourtwov}{Do vulnerabilities in \util files have a high likelihood of being repeated?}

\newcommand{\rqfourthree}{RQ 4.3 Vulnerability Types}
\newcommand{\rqfourthreev}{What are the types of vulnerabilities that appear in \util files?}

\newcommand{\projectYears}{147 }
\newcommand{\totalMeasurements}{1773 }

\definecolor{custom-gray}{cmyk}{0, 0, 0, 0.7, 1.00}
\usepackage[most]{tcolorbox}
\usepackage{cleveref}
\newtcbtheorem[no counter]{Summary}{\hskip-0.97em}{enhanced,drop shadow={black!50!white},
 coltitle=white,
 top=0.15in,
 separator sign none,
 attach boxed title to top left=
 {xshift=0em,yshift=-\tcboxedtitleheight/2},
 boxed title style={size=small,colback=custom-gray}
}{summary}

\begin{document}


\title{Unsafe and Unused? A History of Utility Code in Mature Open Source Projects}

\author{Brandon Keller}

\affiliation{%
  \institution{Rochester Institute of Technology}
  \city{Rochester}
  \state{New York}
  \country{USA}
}
\orcid{0000-0002-4271-9318}
\email{bnk5096@rit.edu}

\author{Kaitlin Yandik}
\affiliation{%
  \institution{Rochester Institute of Technology}
  \city{Rochester}
  \state{New York}
  \country{USA}
}
\email{kmy1691@rit.edu}
\orcid{0009-0006-8569-003X}

\author{Angela Ngo}
\affiliation{%
  \institution{Rochester Institute of Technology}
  \city{Rochester}
  \state{New York}
  \country{USA}
}
\email{an9323@rit.edu}
\orcid{0009-0001-6309-103X}

\author{Andy Meneely}
\affiliation{%
  \institution{Rochester Institute of Technology}
  \city{Rochester}
  \state{New York}
  \country{USA}
}
\email{axmvse@rit.edu}
\orcid{0000-0002-4850-1408}

\renewcommand{\shortauthors}{Keller et al.}

\begin{abstract}
Filenames are a concise means of conveying information about source code to fellow developers. 
One such convention is \utiln. 
Commonly understood to stand for ``utility'', filenames with the letters \util are often an indication that the file contains code that may be broadly useful or reusable. 
Some projects use this convention heavily, for example, the Apache Tomcat server contains 925 files with \util in the path name, which is 17.9\% of all source code files in the tree.  
While the intent of the name may be to prevent duplicate code and reduce workload, what actually happens to \util code over time? 
Do projects move away from \util code as they mature? 
Are \util files being used by fellow colleagues, or maintained and used by their author? 
The goal of our work is \goal. We conducted a longitudinal mining study of the Git repositories of seven open source projects that have a long development history (Linux kernel, Django, FFmpeg, httpd, Struts, systemd, Tomcat).  We analyzed how \util usage, complexity, developer collaboration, and security are potentially correlated within these projects. Our longitudinal analysis was measured at 30-day intervals throughout the entire history of each project, resulting in \totalMeasurements snapshots over \projectYears project-years of development. We conducted rename tracking at every 30-day snapshot to examine \util files over their entire lifetime in a codebase. For example, we found that a \util file can be as much as 2.75 times more likely to be involved in a vulnerability than non-\util files. While every project can adopt their own naming conventions, the ubiquity and longevity of \util files shows a broader developer intent that is useful for understanding the socio-technical nature of software development.
\end{abstract}



\received{20 February 2007}
\received[revised]{12 March 2009}
\received[accepted]{5 June 2009}

\maketitle

\section{Introduction} \label{Introduction}
Utility files often take a unique place in software projects as a home for centralizing functions and methods that contain commonly used functionality, or are otherwise useful outside of the originally intended context, including for other developers working on the project. While these files are highly varied in use and functionality, they are often identified by a common shorthand, \utiln, appearing in a file's path. For example, the Linux kernel has a file called \texttt{ipc/util.c} that provides a variety of useful functions for any developers working on inter-process communication.

There are many benefits to this naming convention. The use of \util files can flatten the learning curve associated with joining a project as complex operations can be in some respects abstracted away to a function or method call reducing the prerequisite knowledge a developer needs to have before making their contributions. In general, workloads can also be reduced as the need to write new implementations of common functionality can be eliminated. This includes benefits in regard to the project's security. Rewriting operations can increase the risk of a software weakness being introduced, as code complexity is a strong predictor of vulnerabilities \cite{shin_empirical_2008} and therefore placing these repeated operations in \util files and reducing rewrites to a single function call can reduce the risk for a vulnerability being introduced in these files and modules making use of \util files. 

These benefits constitute the incentive in the open source context for projects to make use of \util files; however, with the concentration of reusable code in \util files comes the risk that \util files become highly complex and therefore more likely to contain vulnerabilities.  

The goal of this work is \goal. We mined the Git repositories of seven prominent and mature open source projects as a set of case studies representatives of successful long-term software engineering efforts. We collected usage and complexity metrics over time for each project to identify the properties of their \util and non-\util files. We also incorporated analysis of vulnerability to examine the security implications of \util files and the metrics correlated with their historical security. We address the following research questions:

\begin{itemize}
    \item \textbf{\rqone}\footnote{To avoid repetition, all RQs are scoped to mature open source projects} \\ \rqonev
        \begin{itemize}
            \item \textbf{\rqoneone}. \rqoneonev
            \item \textbf{\rqonetwo}. \rqonetwov
            \item \textbf{\rqonethree}. \rqonethreev
        \end{itemize}
        
    \item \textbf{\rqtwo} \\ \rqtwov
    \item \textbf{\rqthree} \\ \rqthreev
    \item \textbf{\rqfour} \\ \rqfourv
    \begin{itemize}
        \item \textbf{\rqfourone.} \rqfouronev
        \item \textbf{\rqfourtwo.} \rqfourtwov
        \item \textbf{\rqfourthree.} \rqfourthreev
    \end{itemize}
    
\end{itemize}

The rest of this paper is organized as follows: Section \ref{Background} details necessary background information and provides definitions for our terminology. Section \ref{Related} discusses related works and their relevance to our study. Section \ref{Methodology} details the methodology we used in answering our research questions. Section \ref{Results} explores the results of our analysis. Section \ref{Validity} discusses the threats to our study's validity. Section \ref{Discussion} presents an expanded discussion on the \util naming convention. Section \ref{Summary} provides our closing remarks and a summary of our contributions.

\section{Background \& Terminology} \label{Background}
In this section, we provide definitions and background information on the terms used in this paper.

\subsection{Common Vulnerabilities and Exposures (CVE)}
The CVE \cite{noauthor_cve_nodate} is a worldwide project for cataloging vulnerabilities in software. CVE data is available from multiple sources including NIST's NVD \cite{booth_national_2015} and MITRE's cve.org \cite{noauthor_cve_nodate}. Not all CVE vulnerabilities have been exploited in the wild. However, being an entry in the CVE is an indication of a significant enough security problem that the development team wanted to announce it to their stakeholders. It is important to note that CVEs lag behind reality as vulnerabilities can exist for long periods of time prior to discovery and can be further delayed by disclosure processes \cite{mcqueen_empirical_2009}. Additionally, the databases containing CVE records do not represent a complete collection of all vulnerabilities found within a project.

\subsection{Common Weakness Enumeration (CWE)}
The CWE \cite{noauthor_cwe_nodate} is a taxonomy of software weaknesses that can manifest as vulnerabilities. The CWE is maintained and made available by MITRE. It is commonly used for CVE and, more generally, vulnerability classification. The CWE is organized as a hierarchy consisting of \textit{Pillars}, \textit{Classes}, \textit{Bases}, and \textit{Variants}. Pillars are the most broad and group together Classes, Bases, and Variants by general themes. Classes are more specific than Pillars and group together other Classes, Bases, and Variants by common properties. Bases represent weaknesses with specific enough information that mitigation strategies and detection techniques are discussed in their entries. Variants are more specific, detailing weaknesses linked to specific technologies or languages with more detail than provided by a Base. While the hierarchy often is organized with Pillars being above Classes, Classes above Bases, and Bases above Variants, this is not always the case and many CWE entries break this pattern. 

\section{Related Work} \label{Related}
Related work spans a variety of topics including utility files generally, software naming conventions, API comprehension, and associations between complexity and vulnerabilities.

\subsection{\util Files}

While mention of \util files has appeared in a few academic research publications \cite{Arakawa2022, Dig2009, Sieck2021, Avidan2017}, including security venues \cite{Sieck2021}, we know of no research that has been conducted in directly investigating the \util convention empirically. The closest research to this is in the program comprehension field that covers naming conventions. For example, Avidan et al. \cite{Avidan2017} studied naming conventions of variables within \util files, but \util files themselves were not in scope. To our knowledge, this is the first academic study focused directly on the \util naming convention.

\subsection{Naming Conventions}
Researchers have investigated naming conventions more broadly outside of \util files. Butler et al. \cite{butler_mining_2011} detailed the naming conventions of Java classes and the associated patterns that they revealed through the mining of 60 open source projects. While they discuss the \texttt{Java.util} package, they focused on the naming conventions of the various classes within the package, notably the convention of preserving the name of interfaces in the classes that implement them. 

Alsuhaibani et al. \cite{alsuhaibani_naming_2021} explored the opinions expressed by 1,100 developers in a survey regarding their thoughts on various standards often applied in the naming of methods such as using full words rather than abbreviations and using verbs to outline a method's behavior. The survey results showed that professionals widely found value in clearly described standards for the naming of methods and that they can be useful in improving code comprehension.

Amit \& Feitelson \cite{amit_language_2022} investigated the naming of software elements in the context of natural language. Specifically, they worked to show that the distribution of names in software is similar to that of words in spoken languages. They found that like traditional language, names in software tend to follow a Zipf distribution, meaning that shorter names and words are more common than longer names. Additionally, they found many compound names of multiple common words. We have made similar observations:  many \util files having their names describe the purpose or function of the file in addition to specifying that it is a \util file (e.g. \texttt{file\_util.c}).

\subsection{API Comprehension}

Developers use \util files much like a remote API in open source projects. Recent research has explored how developers comprehend APIs and effectively make use of them. Heinonen \& Fagerholm \cite{heinonen_understanding_2023} interviewed developers to understand their process when it comes to making use of an API. They found that developers, in addition to reading the source code providing the functionality, rely on documentation and other examples of the API's use. In another work, Sunshine et al \cite{sunshine_2015} investigated the common characteristics of developers' questions related to API usage. They found five common characteristics of these questions, and further found that documentation of APIs does not often match what is needed by developers for proper understanding.

\subsection{Complexity \& Vulnerabilities}

Previous work has explored the empirical association between complexity and vulnerability at the file level. Meneely et al \cite{meneelyTSE2011} explored this relationship, in part, with two case studies. We used a similar methodology in analyzing the seven projects explored in this study, with the major difference being the tools used to collect complexity metrics. 

Mashhadi et al \cite{Mashhadi2023} used Large Language Models to predict bug severity at the method level using complexity among other software metrics as inputs for the model. While different from our approach, they found success in using these metrics in combination with the relevant source code to predict the presence of vulnerabilities. Similarly, Suneja et al \cite{Suneja2023} explored the use of various machine learning models in vulnerability detection with complexity-ranked training.

\section{Methodology} \label{Methodology}
We discuss our methodology for choosing the projects of our analysis and our collection of data about \util files. The scripts used for data collection and processing are provided at the link below along with consolidated data. The full data collection process yields up to 350 gigabytes of information when using our provided set of analysis commits. The code and data utilized for this work is available at \\\href{https://anonymous.4open.science/r/Util-Files-F92C/README.md}{https://anonymous.4open.science/r/Util-Files-F92C/README.md}.

\subsection{\util Files}
We define a \util file as any source code file with the substrings \util or \helper in its file path relative to the root of its repository, ignoring case. For example, all of the following are considered \util files in this study: \texttt{util.c}, \texttt{util/log.c}, \texttt{src/CalcUtil.java}, \texttt{helper.c}, \texttt{src/LogHelper.java}, and \texttt{helpers/log.c}.

\subsection{Project Selection}
Since we're looking at the long-term effects of \util files, our priority in data collection is in finding mature, successful, open source projects that have longevity. We also chose these projects to have a diversity of programming languages and application domains, as well as different styles of open source development. In all cases, we use the entire Git history of each project as available on GitHub. These histories capture \projectYears project-years of software development by thousands of developers.

Additionally, since we are examining vulnerabilities, we are choosing a curated dataset where the vulnerability data has had additional validation. While the NVD \cite{booth_national_2015} contains a massive collection of CVE records, many of the records do not contain adequate information to trace the vulnerability reports directly to the code fixed. Since our study involves source code, this traceability is critical. We chose the data from the Vulnerability History Project (VHP) (\texttt{v7.0-e2e26e9752f}) \cite{meneely_vulnerabilityhistoryprojectvulnerabilities_2023} as the basis for this project, as it involves an editorial process in addition to the NVD to validate the existence of the vulnerabilities as well as the original vulnerability-contributing commit.  We explored seven of the eight projects investigated by the VHP. Chromium was excluded from our analysis due to the size of the repository far exceeding the size of any of the others and its inclusion, in the repository, of many of its dependencies. 

Table \ref{tab:projects} shows the projects and vulnerabilities. Note that the ``Languages'' column denotes the languages identified from the vulnerability patches, though the project itself may involve other languages not mentioned here.

\begin{table}[]
\caption{FOSS Projects in dataset and vulnerabilities collected}
\label{tab:projects}
\begin{tabular}{|l|l|c|c|}
\hline
\textbf{Project} & \textbf{Language(s)} & \textbf{Vulnerabilities} & \textbf{Snapshots} \\ \hline
Linux            & C/C++                        & 2,507                 &  243 \\ \hline
FFmpeg           & C/C++                        & 287                   &  288 \\ \hline
Tomcat           & Java                         & 188                   &  227 \\ \hline
httpd            & C/C++                        & 178                   &  336 \\ \hline
Django           & Python                       & 100                   &  238 \\ \hline
Struts           & Java, Javascript             & 52                    &  205 \\ \hline
systemd          & C/C++                        & 32                    &  205 \\ \hline
\end{tabular}
\end{table}




\subsection{Project Evolution via 30-day Snapshots}
\label{sec:evolution}

This subsection details our process for navigating the history of each repository of our analysis.

To develop a picture of how each project has evolved in terms of its use of \util files, we collected our metrics over time for each project. The projects collectively contain over a million commits, each  documenting meaningful change to the project. In Git, each commit represents a snapshot of the source code tree at the time of the change. We determine our \textbf{snapshot commits} for each project by starting at the repository's initial commit, then determining the first commit that is at least 30 days after the previous, until every snapshot commit is determined. We collect our metrics for our research questions at each of these 30-day snapshots, and analyze the results according to our research questions. 

Our choice of 30 days is arbitrary, but we chose it as a natural rhythm, as many projects have a rapid release schedule \cite{mantyla_rapid_2013}. One could repeat this analysis at larger or smaller windows using our replication package. 

A file could be renamed in between snapshot commits, which would throw off various metric counts, so we conducted \textbf{rename tracking} to ensure that a single file's history is not lost. We used Git's built-in rename detection functionality with the following command:

\begin{verbatim}
git log --pretty="" --name-status \
        --find-renames=50 | grep '^R'
\end{verbatim}

 This command fetches the Git-detected renames with at least 50\% lines matching between renames. We construct a dictionary that maps all historical files as identified via \texttt{git log}, to a list containing all relevant aliases. We use this most significantly in RQ 3 when tracking developer collaboration on \util files throughout the project's history across multiple rename events.


\subsection{\rqone} \label{Usage}

Software projects represent the coming together of many developers and their own unique ways of writing code and addressing problems. Among the ways developers communicate with each other, \util files are but one convention. The reality, however, of what \util files are, how common they ought to be, how they should be used compared to other files, and when the naming convention should be applied or removed to a file are specifics that must be decided upon in each project by the developers developing and making use of them. 

\textbf{\rqone.} \textit{\rqonev}

We analyzed these choices through the Git records of these projects and evaluated the projects' use of \util files in three ways: prevalence, invocation, and renaming.

\subsubsection{\rqoneone}

\textit{\rqoneonev}\\
The first way in which we evaluated usage was in determining the prevalence of \util files in each project, that is, what percentage of the source code files present in the project are \util files. We consider a file to be a source code file if its extension is listed in the VHP repository as a source code file for the given project. These extensions represent files that contain source code at the exclusion of those files in the project that contain no code or development data such as images.

At each snapshot, we used Python's \texttt{os.walk} function to navigate the entire repository. For each file, if its extension was valid based on the VHP list, we evaluate if it is a \util file, a \test file, both, or neither and count up the total of encountered files for each time. 

\subsubsection{\rqonetwo}

\textit{\rqonetwov}\\
The second way we analyzed the usage of \util files was by measuring the usage of functions and methods in \util files compared to non-\util files. We performed a static analysis of the source code files in the repository at each snapshot. 

We began our analysis for each snapshot by running Universal Ctags \cite{Ctags-repo} and generating a JSON output file for each project. Universal Ctags is a currently maintained version of Ctags, a program designed to identify key elements in a provided source code file. For our purposes it is used to identify function and method definitions and the line numbers they can be found on. We use this Ctags data to build a file hierarchy in the form of a tree that contains everything from the root directory to the functions belonging to each file. In addition to this tree structure, we further constructed a dictionary that maps the name of all detected functions to a list of the tree nodes containing a function of the same name. 

We then perform an automated analysis of each source code file in the repository at the time of the snapshot. We used the regex \texttt{\textbackslash b(\textbackslash w+\textbackslash d*)\textbackslash s*\textbackslash (} to identify function calls and then, using the tree previously constructed, attempt to identify the file and function or method being accessed with the function or method call. We start by checking if the function being invoked belongs to the current file itself. Then we check if the function can be found in the other files at the same level of the directory hierarchy. If a match is not found in these two checks, we then see if the function name is unique. If the function's name is not unique and we could not identify a match based on our search on the tree, we opt to over count calls and consider the call being made to be a call to all instances of the function found in the project. In all cases, we record the file(s) being called, the function being called, the function where the call is being made, and the file containing the call. 

To reduce the number of apparent duplicate calls and to reduce the complexity of the tree structure used in our analysis, we excluded both \test and \textit{.h} files from our consideration for the purposes of this research question.

\subsubsection{\rqonethree}

\textit{\rqonethreev}\\
Using our rename tracking methodology, we count adoptions, abandons, and oscillations. An \textbf{adoption} means that a file was renamed from non-\util to a \util file and never went back. Likewise, an \textbf{abandon} is when a \util file was renamed to a non-\util file and never went back. An \textbf{oscillation} is when a file has experienced both a rename from and to \util after multiple renames. 

\subsection{\rqtwo}

The complexity of a file illuminates, in part, its perceived maintainability \cite{antinyan_evaluating_2017}, and in the case of code meant to be used by others, the amount of effort that can be saved by using the preexisting functionality instead of writing a new implementation. If \util files are functioning as hubs for common functionality, we hypothesize them to be more complex.

\textbf{\rqtwo.} \textit{\rqtwov}

We measured the estimated complexity of \util and non-\util files alike across the history of each project. To accomplish this analysis we made use of the Sloc, Cloc, and Code (scc) \cite{boyter_boyterscc_2025} tool on every snapshot for each project. This tool supports every prominent language used in the projects we have analyzed. All analysis was conducted with scc version 3.4.0. 

We used the \texttt{by-file} option in scc to produce results on a file-by-file basis instead of as combined results for each language. This was done to allow for comparison between \util and non-\util files. Additionally, this mode allows for the optional exclusion of \test files from our data. 

We write all of our acquired data into a CSV file where each line includes the language the file was written in, the path to the file, the file's name, and the full set of complexity and line number measurements provided by scc.

We normalized all complexity data by source lines of code (SLOC). This SLOC measurement was determined alongside estimated complexity by the scc tool. In addition to using this data for normalizing the reported complexity measurements, we further used these SLOC results as a means to evaluate the size of the project over time. 

\subsection{\rqthree}

\label{sec:method-collaboration}

Developer communities change over time as new developers join a project and others change what they work on. While a function may have a simple start with a single developer writing it to call in their own work later on, in large open source projects it is possible for even those functions with simple origins to become the target of work for a great many developers over the course of many years. As functions receive development, they may also prove useful to other developers that may come to make use of its functionality. We sought to understand the evolution of \util file development communities through an analysis of \texttt{git blame} records. Our research question is therefore: 

\textbf{\rqthree.} \textit{\rqthreev}

We performed \texttt{git blame} operations on every source code file in the repository and saved this data as a JSON file where every line in each file is mapped to the author of the line and the time of the commit that resulted in the line appearing in the snapshot. As we progressed through the snapshots, we used \texttt{git diff} operations to identify files that had been changed since the previous snapshot. We then performed additional \texttt{git blame} operations on those files that changed and updated any relevant JSON data to reflect the most up-to-date information. For the purposes of our processing of data, we read the JSON blame data into a dictionary for simple mapping of a line in a given file to the author that made the most recent change to the line.

The analysis that we perform for the purposes of this research question after collection of \texttt{git blame} data is further reliant upon the data collected and the intermediate structures constructed for RQ 1 as detailed in Section \ref{Usage}, specifically the map of function/method calls to the function or method being called, the Ctags data used for identifying functions, and the identified function calls in all source code files.  

For each function call that we identified across the repository, we accessed the blame dictionary we previously constructed to identify the author. Additionally, we performed this process for every line of every function identified by Ctags for the given snapshot. As Ctags does not report the end line of functions, we consider a function to end one line prior to the next function listed by Ctags. In the case of the final function in a file, we consider all remaining lines in the file to belong to that final function.

To determine the types of collaboration occurring for any given function, we made use of the function call mapping in combination with the Git blame data to group together all developers that have worked on or with a function or method in any given snapshot. We created a data structure to store this information that consisted of a dictionary that mapped files to a dictionary that maps functions to another dictionary that maps authors to a list of data. This list of data was two elements in length. The first index of this list contained the first time that the author had made a modification to the function in question. The second index contains the first time that the author made a call to the function in question.

For each snapshot processed, we updated the dictionary of files to authors and their contribution times to reflect the most recent information. When we encounter a file that has been renamed, as identified in our renaming analysis for RQ 1, we replace the entry with the new file name as the key. To prevent double counting, the duplicated version with the old file name as the key is removed from the dictionary.



Using invocation data and authorship data together, we define the following metrics:

\begin{itemize}
    \item \textbf{Function Author Only (FAO)}: the author has only contributed to the function in question and has not called to it themselves
    \item \textbf{Call Author Only (CAO)}: the author has only made calls to the function in question and has not contributed to the function itself
    \item \textbf{Function Author to Call Author (FTC)}: the author began by contributing the function in question and in a later commit started making use of the function with a call to it
    \item \textbf{Call Author to Function Author (CTF)}: the author used the function via a call and later contributed to the function as an author
    \item \textbf{Same Time Contributor (STC): } the author modified the relevant function in the same commit that they began calling the function elsewhere.
\end{itemize}

At the conclusion of processing for each snapshot, we determined a collaboration type for each author found in the project based on their earliest contributions to the function in question be it a call or a modification to the function code. We then determined, overall for each snapshot for every function what types of collaboration were occurring in the development process.



\subsection{\rqfour}

Mistakes can propagate through a system through function calls and data transfer, and \util files, if widely used, may serve as a conduit for spreading a mistake to otherwise disconnected subsystems. Widely-used functions may also be at the most risk of accidental misuse by other developers thanks to the number of developers making use them. \util files are therefore likely to be at an elevated risk in many projects. We explore the historical security of each project and their \util files to uncover if \util files are truly higher risk.

\textbf{\rqfour.} \textit{\rqfourv}

Our investigation into the security of \util files began by first downloading the complete set of \textit{Offender Files}, which are files that are found, by the VHP, to be involved in the fix commit of a vulnerability. This set of files is available via the VHP API. From this set of offender files, we then again used the VHP API to identify the CVEs associated with each of the offender files. With the CVEs in hand we made an additional set of API requests to the VHP to get the event logs for each of the CVEs. While this log may contain many distinct events in the vulnerability's history, we focus on and store for later processing only the fix commit, and more specifically, the fix commit's date.

With renames and API data in hand, we consolidate our data to a format that is more practical for the processing to be undergone during the navigation through snapshot commits. We created a smaller dictionary for each project to map file paths to the date after which we consider the file to be an offender or previously vulnerable. Each file name appears separately in this dictionary to simplify lookup procedures, but the date of vulnerability is maintained across renames.

\subsubsection{\rqfourone}
\textit{\rqfouronev}\\
We approach the historical security of the analyzed projects from three directions. We first considered the relative risk of \util files compared to non-\util files. With the full set vulnerable files and the dates at which we can classify them as vulnerable we developed a data structure to contain our resultant risk data. For each project we maintain a list where each index represents a snapshot. In each index is another list containing eight elements where the first four are: the number of vulnerable \util files present, the number of non-vulnerable \util files, the number of vulnerable non-\util, and the number of non-vulnerable non-\util files. The final four elements are the same with the exception that \test files have been removed from consideration.

Once all the data had been collected, we then calculated odds ratios for each snapshot indicating the odds of a \util file being vulnerable compared to the odds of a non-\util file being vulnerable. For this calculation we made use of the following equation:

\begin{displaymath}
  \frac{(\frac{\texttt{Offending Util Files}}{\texttt{Non-Offending Util Files}})}{(\frac{\texttt{Offending Non-Util Files}}{\texttt{Non-Offending Non-Util Files}})}
\end{displaymath}

We further performed these calculations for the data with \test files excluded.

\subsubsection{\rqfourtwo}
\textit{\rqfourtwov}\\
In our second approach we attempted to measure the security of each project through \textit{Recidivism Metrics}. We consider two recidivism metrics for this application, they are detailed below:

\paragraph{Module Recidivism:} The number of vulnerabilities fixed during a snapshot that involve fixes to previously fixed files or modules\\

\paragraph{Type Recidivism:} The recurrence of a CWE in a vulnerability after the same CWE has already been corrected as part of another vulnerability in the same project\\

These metrics aim to describe the health of a project's process in terms of adapting to security situations without the need to rely only on the number of vulnerabilities that have been introduced in the project. A high rate of recidivism, for example, would indicate that a project team's process is not facilitating the learning required to prevent recurrent problems from occurring. 
Each of the recidivism metrics has a corresponding equation that we use for calculating a snapshot's recidivism rates. The equation for Module Recidivism is below:

\begin{displaymath}
  \frac{\texttt{CVEs Fixes that updated a fixed file in Last 30 Days}}{\texttt{Total CVEs Fixed in the Last 30 Days}}
\end{displaymath}

The equation for Type Recidivism is below:

\begin{displaymath}
  \frac{\texttt{CVEs Fixed with Recurrent CWE in Last 30 Days}}{\texttt{Total CVEs Fixed in the Last 30 Days}}
\end{displaymath}

These equations are calculated 3 times for each snapshot. First, these metrics are calculated overall including all files. Then these metrics are calculated only inclusive of CVEs with \util files in their fix commit. Lastly, these metrics are calculated only considering CVEs without any \util files in their fix commits. 

In addition to the 30-day time scale we described in the equations above, these metrics are flexible to fit the needs of the project. For our purposes here, we calculate each of these metrics for both the 30 and 90-day time scales. Depending on the amount of CVEs reported in a given period, it may be impractical to rely on the 30-day scale alone. 

\subsubsection{\rqfourthree}
\textit{\rqfourthreev}\\
The third way in which have approached the security of these projects and their \util files is by analyzing the types of vulnerabilities that have been historically present in the projects. For comparison, we explore vulnerabilities that include \util files in their fix separately from those vulnerabilities that do not include \util files in their fix. In all cases, we use the CWE identifications provided by the VHP.


\section{Results} \label{Results}


\begin{table*}[]
\caption{Oscillation, Adoptions, and Abandon data for each of the seven projects.}
\label{tab:promotions}
\centering
\begin{tabular}{|l|c|c|c|c|c|c|c|}
\hline
\textbf{Project} & \textbf{Osc.} & \textbf{Adopt.} & \textbf{Abandon} & \textbf{Adopt./Abandon} & \textbf{Osc. NT} & \textbf{Adopt. NT} & \textbf{Abandon NT} \\ \hline
Django         & 0  & 9  & 7  & 1.29 & 0  & 3  & 1  \\ \hline
FFmpeg         & 0  & 40 & 28 & 1.43 & 0  & 39 & 27 \\ \hline
HTTPD          & 0  & 4  & 1  & 4.00 & 0  & 4  & 1  \\ \hline
Linux Kernel   & 11 & 47 & 78 & 0.60 & 11 & 47 & 70 \\ \hline
Struts         & 0  & 0  & 12 & 0.00 & 0  & 0  & 8  \\ \hline
systemd        & 6  & 56 & 15 & 3.73 & 5  & 27 & 12 \\ \hline
Tomcat         & 0  & 52 & 9  & 5.78 & 0  & 39 & 4  \\ \hline
\end{tabular}
\end{table*}

In this section, we report the results for each research question.

\subsection{\rqone}

\textit{\rqonev}. We examined prevalence, invocation, and renaming.

\subsubsection{\rqoneone}
\textit{\rqoneonev}

\begin{figure}[h]
    \centering
    \includegraphics[width=.9\columnwidth]{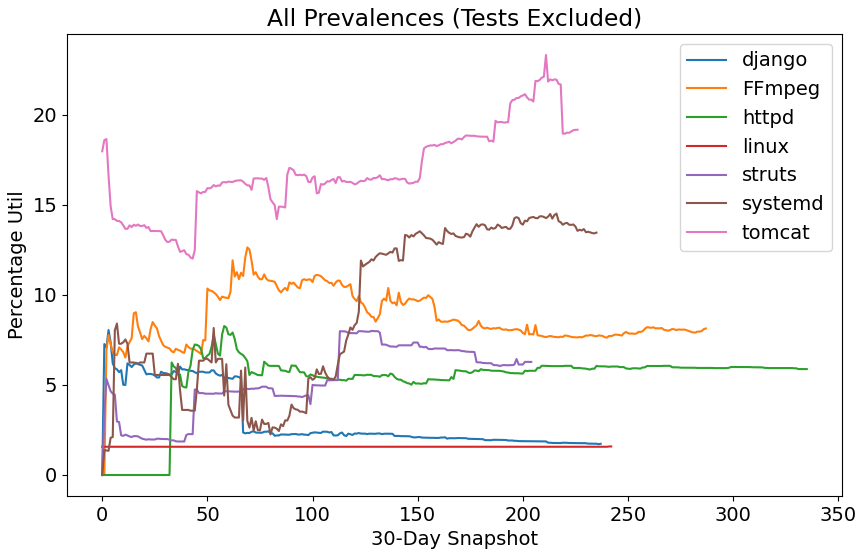} 
    \caption{Prevalence of \util Files across all seven projects, excluding test \util files.}
    \label{fig:prevalence}
\end{figure}

All projects used the \util naming convention, to varying degrees. The Linux Kernel used it most sparingly, with at minimum having 1.5\% of non-test source code  files being \util. This still amounts to 1,136 source code files, across the Linux Kernel codebase at that time. Similarly, Django, which consistently has also had among the lowest ratio of \util files to other source code files maintains a code base of 1.7\% \util files. Worth noting, however, is that the Linux Kernel has maintained a consistent \util prevalence since its first snapshot while Django has had over 8\% of its source code files as \util files in its earliest snapshots.

Outside of Linux and Django, the percentages of \util files making up each project repository, as of the most current snapshots with tests excluded, range from 5.8\% of files being \util in Apache HTTPD to over 19\% of files being \util in Apache Tomcat. A graph of \util file prevalence is shown in Figure \ref{fig:prevalence}. We observe an overall upward trend in \util usage over the life of projects, with periods of decreased usage. 

\begin{Summary}{Prevalence}
    X \util files are commonplace, making up to 23.3\% of files.
\end{Summary}

\subsubsection{\rqonetwo}
\textit{\rqonetwov}



Each of the seven projects has, for at least the 50 most recent snapshots, had at least the same amount of average calls to \util files compared to the average calls to non-\util files. The Linux kernel has historically had very similar usage between \util and non-\util files, but the other projects have had higher rates of \util usage ranging up to seven times as often as non-\util files. 

Django and HTTPD have maintained the highest ratio of average \util calls to average non-\util calls for over 100 snapshots worth of development. Both of these projects also have among the lowest percentage of their files adhering to the \util naming convention with only Linux having fewer \util files over the most recent years of development. In contrast, Apache Tomcat has maintained an average call ratio only between one and two, and yet ranks the highest in terms of the percentage of its files adhering to the convention. 

\begin{figure}[h]
    \centering
    \includegraphics[width=.9\columnwidth]{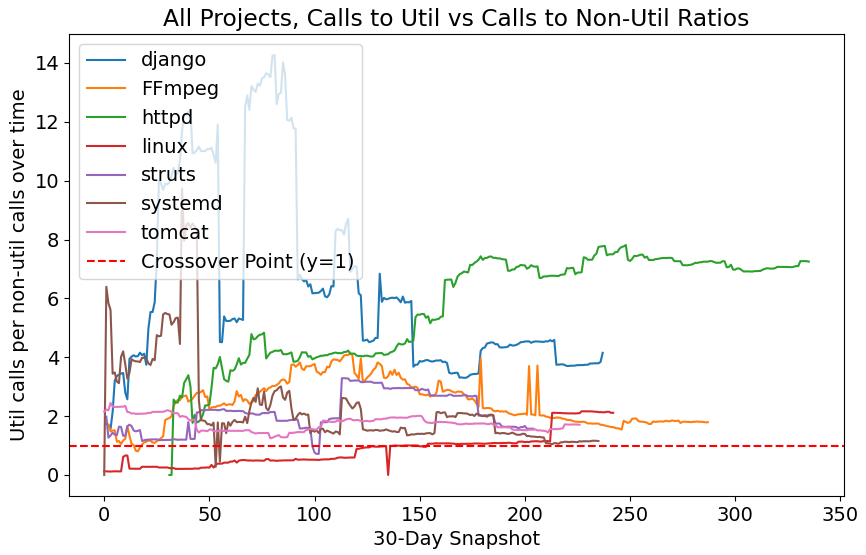} 
    \caption{Usage Ratios of \util functions to non-\util functions}
    \label{fig:usage}
\end{figure}

Figure \ref{fig:usage} shows the historical average usage of \util files compared to the average usage of non-\util files for each of the projects' entire set of snapshots. 

\begin{Summary}{Invocation}
    X \util functions are used up to 7 times more than other functions.
\end{Summary}

\subsubsection{\rqonethree}
\textit{\rqonethreev}

We found 375 unique files across the seven projects analyzed that experienced adoptions, abandons, or oscillations and that for five of the seven projects, for every one abandon there were at least 1.2 adoptions. Adoptions of non-\util files into the \util naming convention were the most common with a total of 208 Adoptions. Abandons were the second most common with a total of 150 events. Oscillations were the least common with only 17 total instances in the seven projects, all of which came from the Linux Kernel or systemd. With tests included, adoptions were 38\% more common that abandons and with tests excluded, 29\% more common. A table detailing which projects included which events can be found in Table \ref{tab:promotions}.

Again, the Linux Kernel differs from the other projects in experiencing more abandons than adoptions with 78 abandons compared to only 47 adoptions. In addition, 10 of the 11 oscillating files belonging to the Linux Kernel repository currently remain as non-\util files. Of the six oscillations in the systemd repository, two end as \util files. 

\begin{Summary}{Renames}
    X Files both adopted and abandoned the \util convention. 
\end{Summary}

\subsection{\rqtwo}
\rqtwov

\begin{figure}[h]
    \centering
    \includegraphics[width=.9\columnwidth]{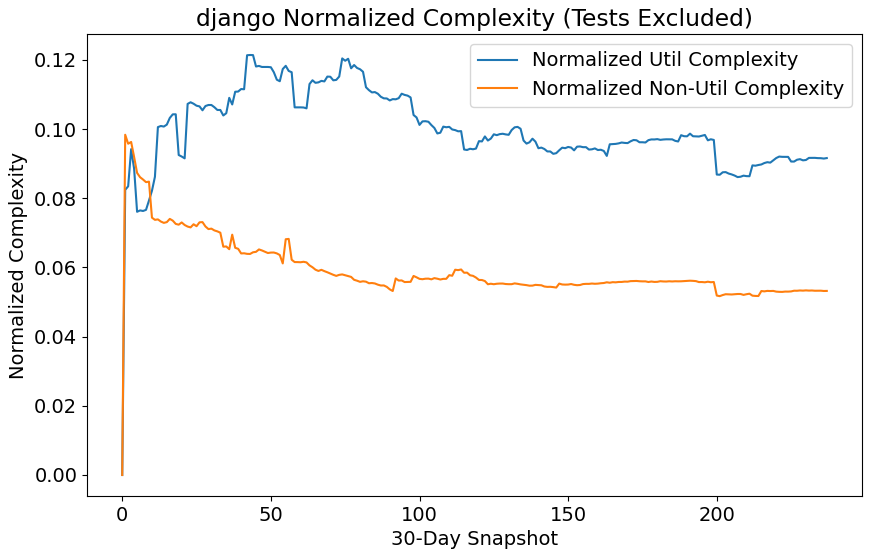} 
    \caption{Django Complexity Data. Test Files Excluded}
    \label{fig:complexity}
\end{figure}

We found that six of the seven projects consistently had, on the whole, \util files more complex than non-\util files. For at least the most recent 50 snapshots, they have had normalized \util complexity values over .015 greater than their normalized non-\util complexity values. FFmpeg was the exception to this finding as \util files were instead less complex or just as complex as non-\util file as over the course of the project's evolution. Within the most recent 100 snapshots, \util complexity in the FFmpeg project has also decreased. Apache Tomcat, systemd, Apache HTTPD, and the Linux Kernel have not, in the analyzed snapshot ranges, ever experienced a period while \util files were in the project that \util files were not overall more complex than non-\util files. In the remaining projects where periods of higher non-\util complexity occurred, these periods always occurred within the first half of the project's snapshot set.



As an example, the normalized complexity data for Django is shown in Figure \ref{fig:complexity}.

\begin{Summary}{Complexity}
     X Six of seven projects have \util files that are more complex than other files in recent snapshots.
\end{Summary}

\subsection{\rqthree} \label{CollabResults}
\textit{\rqthreev}

\begin{figure}[h]
    \centering
    \includegraphics[width=.9\columnwidth]{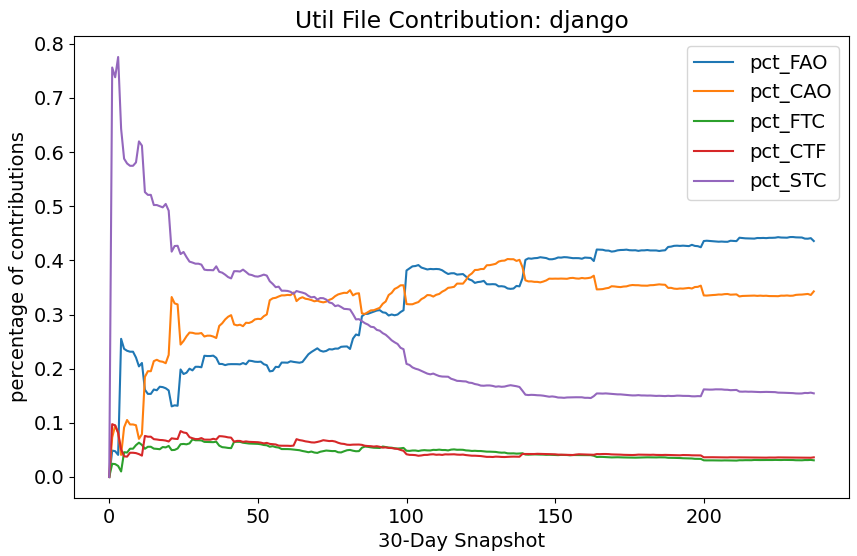} 
    \caption{Collaboration Patterns in the Django Project}
    \label{fig:collab}
\end{figure}

The metrics we defined in Section \ref{sec:method-collaboration} for this question are: Function Author Only (FAO), Call Author Only (CAO), Function Author to Call Author (FTC), Call Author to Function Author (CTF), and Same Time Contributor (STC). 

We found that each of the projects varied greatly in terms who was editing and who was calling functions in \util files. The most consistent result is that, for all projects studied other than the Linux Kernel, at least a plurality of authors invoking \util functions were also editing the same \util functions in the same snapshot (STC) during the earliest five months of development. 

Also consistent across all projects during the most recent 50 snapshots is that no fewer than 57.7\% of \util function contributors and callers had only contributed (FAO) or only called the same \util function (CAO). Whether only calling a \util function or only contributing to a \util function was more common varied by project. Apache Tomcat, systemd, Apache Struts, and FFmpeg have experienced more Call-only authors (CAO) while the Linux Kernel, Django, and Apache HTTPD have experienced more cases of developers contributing to a function and not calling it themselves (FAO).

All projects also experienced between 4.7\% and 17.5\% of function authors that started using a function and then modified the function later on (CTF) or that started developing a \util function and later using it themselves (FTC). Thus, we found it rare for \util functions to be developed by the authors that make use of them. The evolution of Django's \util collaboration makeup is shown in Figure \ref{fig:collab}.



\begin{Summary}{Collaboration}
    X In the 50 most recent snapshots, over 57.7\% of developers working with \util did not both edit and call the function.
    
\end{Summary}

\subsection{\rqfour} \label{SecResults}
\textit{\rqfourv} We examined risk factors, vulnerability recidivism, and vulnerability types.

\subsubsection{\rqfourone}
\textit{\rqfouronev}

\begin{figure}[h]
    \centering
    \includegraphics[width=.9\columnwidth]{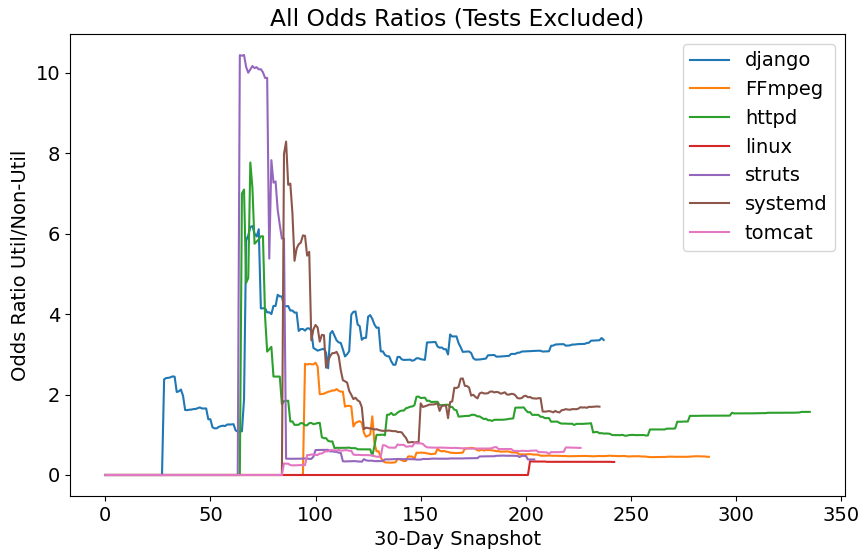} 
    \caption{Evolution of Odds Ratios for each project. Test Files Excluded}
    \label{fig:odds}
\end{figure}

Five of the seven projects investigated had a major spike in their odds ratios during their earliest snapshots when file counts are, in general, much lower than they are now.
For example, Apache Struts reaches odds ratios higher than 10, when excluding test files, during snapshots from 2012 when only six files had been fixed for a vulnerability overall. This means that during these snapshots, a \util file is 10 times more likely to be a part of a vulnerability fix than a non-\util file. In contrast, in 2025, Apache Struts experiences a vulnerability odds ratio of less than 0.5, when excluding test files, with 187 files that have been fixed for a vulnerability overall. The Linux Kernel and Apache Tomcat both do not have any noticeable spikes in their odds ratios in their snapshot history. These projects, however, have a history that predates the Git repository we analyzed. These projects possibly experienced a spike in their \util vulnerability odds ratios prior to their commit history's beginning in Git. 

A graph showing the evolution of odds ratios for each project is in Figure \ref{fig:odds}. Every project, with the exception of Apache Tomcat and the Linux Kernel, experiences a considerable spike in odds ratios at around the same time in their development cycle, between month 50 and month 100 of development. In these spikes, each of the five projects experienced at least a 2.75 odds ratio. It is further worth noting that these spikes, with the exception of Apache Tomcat, Django, and the Linux Kernel, align with the first vulnerability fixes involving \util files. 

\begin{Summary}{Vulnerability Risk}
    X Five of seven projects experienced a vulnerability odds ratio of at least 2.75
\end{Summary}

\subsubsection{\rqfourtwo}
\textit{\rqfourtwov}
Across the seven projects analyzed, 717 unique files were fixed for more than one vulnerability. Of these 717 files, 69 or 9.6\% of them were \util files. Additionally, on a project by project basis, vulnerabilities involving a fix to \util files were recidivistic over 50\% of the time for projects other than the Linux Kernel and systemd. 

In contrast, type recidivism was not observed at rates exceeding 44\% in any project except for the Linux Kernel where 68.4\% of non-\util related vulnerabilities were of the same CWE classification as a previously fixed vulnerability. Tables \ref{tab:recidivism_util} and \ref{tab:recidivism_nonutil} show full recidivism rates for each project across \util related and non-\util related vulnerabilities respectively. 

\begin{table}[]
\caption{The percent of \util CVEs that resulted in recidivism. Percentages rounded to the nearest 10th of a percent.}
\label{tab:recidivism_util}
\begin{tabular}{l|c|c|}
\textbf{Project} & \textbf{\% Module} & \textbf{\% Type} \\ \hline
Django         & 70.8\% & 16.7\% \\ \hline
FFmpeg         & 52.9\% & 5.9\%  \\ \hline
HTTPD          & 76\%   & 16\%   \\ \hline
Linux Kernel   & 0\%    & 0\%    \\ \hline
Struts         & 81.3\% & 43.8\% \\ \hline
systemd        & 14.3\% & 0\%    \\ \hline
Tomcat         & 71.4\% & 28.6\% \\ \hline
\end{tabular}
\end{table}

\begin{table}[]
\caption{The percent of non-\util CVEs that resulted in recidivism. Percentages rounded to the nearest 10th of a percent.}
\label{tab:recidivism_nonutil}
\begin{tabular}{l|c|c|}
\textbf{Project} & \textbf{\% Module} & \textbf{\% Type} \\ \hline
Django         & 66.1\% & 26.8\% \\ \hline
FFmpeg         & 44.2\% & 12.1\% \\ \hline
HTTPD          & 54.9\% & 18.7\% \\ \hline
Linux Kernel   & 25.9\% & 68.4\% \\ \hline
Struts         & 66.7\% & 30.3\% \\ \hline
systemd        & 26.1\% & 8.7\%  \\ \hline
Tomcat         & 88.1\% & 19.0\% \\ \hline
\end{tabular}
\end{table}

\begin{Summary}{Vulnerability Recidivism}
    X 9.6\% of files that were fixed for more than one vulnerability were \util files.
\end{Summary}

\subsubsection{\rqfourthree}
\textit{\rqfourthreev}
Among the seven projects analyzed, there were a total of 3,344 vulnerabilities. 98 of these resulted in a fix to a \util file. Of the vulnerabilities impacting the \util files, the most common types of vulnerability were:
\begin{itemize}
    \item CWE-20: Improper Input Validation
    \item CWE-94: Improper Control of Generation of Code (Code Injection)
    \item CWE-150: Improper Neutralization of Escape, Meta, or Control Sequences
\end{itemize}
Of the 98 \util vulnerabilities, these three were the CWE for 14, 5, and 3 of them respectively. In the context of all CVEs there were 168 occurrences of CWE-20 and 10 occurrences of CWE-94. CWE-150 was uniquely associated with vulnerabilities impacting \util files and therefore had only 3 occurrences in the overall data set.

\begin{Summary}{Vulnerability Type}
    X The most common vulnerability type affecting \util files was CWE-20 Improper Input Validation.
\end{Summary}

\section{Threats to Validity} \label{Validity}

The projects we have selected for our analysis are mature and well maintained by experienced teams and represent projects with resources for code review, testing, etc. that other teams may not have access to. While this set of projects does not represent all open source communities, the fact that these projects are still prone to having dangerous \util files despite their many advantages suggest that all projects could benefit from enhanced care when dealing with \util and other shared files. Additionally, the projects we have analyzed are the projects investigated as part of the VHP. They include curated records for the vulnerabilities we have investigated here. In contrast, large data sets such as the NVD often do not include relevant commit information for identifying the fix or source files that were a part of the fix commit. All of our scripts are made available and can be adapted to support any project provided the fix commit data is provided for any future researchers to explore a wider set of works.

While ``util'' and ``helper'' were common in paths in our set of case studies, they do not represent an exhaustive list of conventions to represent shared code. Other potentially widespread conventions include ``shared'' or ``global''. Our analysis of these projects did not suggest these conventions were prominent, but outside of these projects the trends and patterns we have observed may or may not be shared with projects making use of other conventions.

For the purposes of our analysis, we made use of Git's built-in rename detection. This operation is not perfect and relies on file similarity between name changes. This process leaves the possibility that some renames may not be recognized and that in other cases some files may be improperly recognized as a rename when they are in fact not a part of a rename. Additionally, Git logs may not reflect all changes in a projects history due to rebasing and other Git operations that may impact the forensic soundness of the log. We provide all of our data and all of our scripts used for collecting our data for other researchers to rerun our analysis as needed.

Throughout this work, we consider a file to be vulnerable or an offender if it is included in a vulnerability's fix commit. This was done for two primary reasons. First, the fix commit details are more congruent with the requirements of our recidivism metrics as recidivism is reliant upon the team being aware of the vulnerability and having had the experience in correcting the vulnerability. Second, while initial commits can be accessed through algorithms such as SZZ \cite{sliwerski_when_2005}, there exists a myriad of concerns as to the accuracy of these algorithms \cite{williams_szz_2008, neto_impact_2018, da_costa_framework_2017} in the literature. Other than our recidivism metrics, there is no hard requirement that our tooling be ran exclusively on fix commit data. With our provided data and scripts researchers can explore projects through the lens of vulnerability contributing commits as needed. 

We use light-weight static analysis of source code files to approximate the trace from a function or method call to the function or method being called. While more advanced techniques could yield a more technically accurate measurement of calls to functions and methods, our intention is less centered upon understanding the true usage of \util files, but instead with uncovering the differences in use between \util and non-\util files that can be effectively analyzed and discussed from our simplified analysis. All of our data is available for any comparisons to more formal analysis methods.

\section{Discussion} \label{Discussion}
Our results show that utility files differ between projects and across time with implications for developers more broadly.

\subsection{Linux Kernel Divergences}
Throughout our study, we found that the Linux kernel was consistently the exception to the trends observed in the other projects. Specifically, the Linux Kernel:

\begin{itemize}
    \item had more calls to non-\util files than calls to \util files,
    \item had an adoptions-abandon ratio of less than one, and
    \item did not have vulnerability odds ratios in excess of one.
\end{itemize}

Regarding security, the Linux kernel has only had a single \util-related vulnerability reported in the VHP dataset. For the non-security metrics, the divergence from the observed trends in other projects may be due to the stability of the project in terms of the amount of files being added over time. The Linux Kernel's first snapshot in the Git history we analyzed began with 69,142 source code files and the most recent snapshot ended with 69,536 source code files. This is in contrast to all of the other projects we analyzed, that grew more considerably in number of files over time. Additionally, the Git records we accessed for the Linux Kernel began in April of 2005. This is over 10 years into the development of the Linux Kernel and thus the observations we make about many of the early developments for each of the projects we studied may not hold for the Linux Kernel as those same development milestones happened years prior to the first snapshot we took measurements at.

\subsection{Advice for Maintainers}

From this study, we advise the following to project maintainers.

\textbf{On Meaning}. While the \util naming convention is very widely adopted, every team uses \util a little bit differently. Every project develops their own shorthand language for what a module is to be used for. New developers need to be aware of how the team uses this convention, and to make sure they contribute appropriately. Especially if teams tend to adopt \util more often than abandon it, the intended meaning of the naming convention (and others like it) ought to be documented and conveyed.  

\textbf{On Complexity and Security. } The name of a file does not directly cause issues such as complexity and vulnerability. But, the naming of a file conveys developer intent and is part of intra-team communication. Teams can easily mine their own repository structure and examine if these trends of complexity and vulnerability apply to their code, and can take any additional fortification efforts.

\section{Conclusion} \label{Summary}

The goal of this work is \goal. We found that \util files are more likely to have a history of vulnerability in the majority of projects that we analyzed and that in the same data set, \util files are up to 10 times more likely to be vulnerable at the earliest phases of a project's development when the number of source code files is minimal. We further found that functions in \util files are more used than functions in non-\util files, but that \util code can also be more complex than that of other files. Overall, our results show the promises and perils of highly used \util files throughout their development from single source code files to twenty or more years into development with thousands of contributors. 

\section{Acknowledgments}
This work is supported by NSF Grants 2336252 and 1922169.

\section{Data Availability}
The replication package for this work is made available at \\https://doi.org/10.5281/zenodo.19919905 \cite{brandon_keller_bnk5096util-files_2026}.

\bibliographystyle{acm}
\bibliography{bibfile}

\end{document}